\documentclass[prd,preprint,preprintnumbers,nofootinbib,eqsecnum,superscriptaddress]{revtex4}

 \usepackage[dvips,final]{graphicx}
  \usepackage{amssymb}
   \usepackage{amsmath}
    \usepackage{amsfonts}
     \usepackage{epsfig}
      \usepackage{bm}% bold math

\usepackage{mathpazo}

\usepackage[section]{placeins}

\usepackage{multirow}
\usepackage{ctable}
\usepackage{booktabs}
\usepackage{array}
\usepackage{tabularx}
\usepackage{xcolor}
\usepackage{pstricks}
\definecolor{fred}{rgb}{0.90053, 0.00369, 0.00159}  % ta3skyblue

\newcommand{\Jpsi}{J\!/\!\psi}

\begin{document}

%\vfill
\title{\bm{$J/\psi$}-meson production within improved color evaporation
  model\\ with the \bm{$k_{T}$}-factorization approach for \bm{$c\bar{c}$} production}

\author{Rafa{\l} Maciu{\l}a}
\email{rafal.maciula@ifj.edu.pl} \affiliation{Institute of Nuclear
Physics, Polish Academy of Sciences, Radzikowskiego 152, PL-31-342 Krak{\'o}w, Poland}

\author{Antoni Szczurek\footnote{also at University of Rzesz\'ow, PL-35-959 Rzesz\'ow, Poland}}
\email{antoni.szczurek@ifj.edu.pl} \affiliation{Institute of Nuclear
Physics, Polish Academy of Sciences, Radzikowskiego 152, PL-31-342 Krak{\'o}w, Poland}

\author{Anna Cisek}
\email{acisek@univ.rzeszow.pl} \affiliation{Faculty of Mathematics and Natural Sciences,
University of Rzesz\'ow, ul. Pigonia 1, PL-35-310 Rzesz\'ow, Poland}
%\affiliation{University of Rzesz\'ow, PL-35-959 Rzesz\'ow, Poland}

\begin{abstract}
We use a new approach to color evaporation model (CEM)
for quarkonium production.
The production of $c\bar c$ pairs is performed within 
$k_T$-factorization approach using different unintegrated gluon
distribution functions (UGDF) from the literature. 
%In the standard approach there is only one 
%parameter -- the normalization of the cross section (probability to 
%form the $J/\psi$ meson). 
We include all recent improvements to color evaporation model.
We cannot describe simultaneously mid and forward rapidity data measured
at the LHC when using the KMR UGDF with the same normalization parameter. 
Furthermore we get somewhat too hard distribution in 
$\Jpsi$ transverse momentum.
Correcting the standard KMR distributions for saturation effects at small
values of $x$ improves $\Jpsi$ rapidity distributions, while taking 
correction for emissions hidden in the KMR UGDF in the evaluation of 
$x$ values improves a bit $\Jpsi$ transverse momentum distributions.
%We show that at large rapidities and large transverse momenta (much) bigger
%$| \vec{p}_{diff}| = |\vec p_c - \vec p_{\bar c}|$ are involved.
%It seems physically motivated that the probability depends on $p_{diff}$.
%We suggest to add an extra probability function depending on $p_{diff}$.
%A parameter of the phenomenological function can be then adjusted
%to the LHC data. We get very good fit of the all LHC data with the two
%free parameters.
The modifications improve also description of the LHCb $D$-meson data.
\end{abstract}

\maketitle

%----------------------------
\section{Introduction}
%----------------------------

Inclusive production of quarkonia is one of the most actively studied
topics at the LHC. The $\Jpsi$, $\Psi'$, $\Upsilon$, $\Upsilon'$
and $\Upsilon''$ are the usually measured quarkonia.
The production of $\Jpsi$ is a model case. There was (still is) a
disagreement related to the underlying production mechanism.
There are essentially two approaches. The first one is the so-called
nonrelativistic QCD (NRQCD) approach 
\cite{Chang:1979nn,Berger:1980ni,Baier:1981uk,Braaten:1994vv}. 
There are two versions of such an approach based on collinear 
or $k_T$-factorization approaches. 
It was shown recently that the LHC data can be explained
within the NRQCD $k_T$-factorization approach with a reasonable set of
parameters \cite{CS2018}. 

Another popular approach is color evaporation model \cite{Fritsch,Halzen}.
In this approach one is using perturbative calculation of $c \bar c$.
The $c \bar c$-pair by emitting a soft radiation goes to color singlet
state of a given spin and parity. The emission is not explicit in this
approach and everything is contained in a suitable renormalization
of the $c \bar c$ cross section when integrating over certain limits
in the $c \bar c$ invariant mass. It was proposed recently how to improve
the original color evaporation model \cite{MV2016,CV2017,Cheung:2018tvq}.
The original color evaporation model is based on calculation in
collinear approach. The leading-order (LO) approach is known to be not
sufficient to describe the inclusive charm data. The next-to-leading
order (NLO) approach is much better in this respect \cite{Nason:1987xz,Beenakker:1988bq,Cacciari:2012ny}. There is even
next-to-next-to-leading-order (NNLO) approach \cite{Czakon:2008ii} but only for 
inclusive single charm distributions. One needs to calculate correlation
distributions ($c \bar c$ invariant mass) in the context of 
the color evaporation model 
so the NNLO calculation cannot be used in this context.
On the other hand the $k_{T}$-factorization with the KMR unintegrated
distributions turned out to be successful in the inclusive production 
of $D$ mesons \cite{Maciula:2013wg,Maciula:2018iuh} as well as for some correlation
observables \cite{Maciula:2013wg,Karpishkov:2016hnx} at the LHC. 
It seems therefore interesting, and valueable,
to apply the $k_{T}$-factorization approach for $c \bar c$ production 
in the context of applying the color evaporation model for $\Jpsi$ meson production.

In the present paper we wish to study whether such a combination
of elements can allow to describe the world data for 
$\Jpsi$ production\footnote{When our analysis was almost
finished we found a new prepint \cite{Cheung:2018tvq} which discusses exactly 
the same process. The main difference is a use of different unintegrated gluon distributions.}.

%--------------------------------------------------------------
\section{Theoretical framework}
%--------------------------------------------------------------

In the basic step of our approach, i.e. calculation of the cross section for $c\bar c$-pair production, we follow the $k_{T}$-factorization approach. This framework was shown many times by different authors to provide very good description of heavy quark production in proton-proton collisions at different energies. Some time ago it was successfully used for theoretical studies of $pp \to c \bar c \;\! X$ reaction at the LHC, including open charm meson \cite{Maciula:2013wg,Karpishkov:2016hnx}, as well as $\Lambda_{c}$ baryon production \cite{Maciula:2018iuh}. Very recently, this approach was also applied \textit{e.g.} for $pp \to c \bar c + \mathrm{jet} \;\! X$ \cite{Maciula:2016kkx}, $pp \to c \bar c + \mathrm{2jets} \;\! X$ \cite{Maciula:2017egq} and $pp \to c \bar c c \bar c \;\! X$ \cite{Maciula:2013kd}.

According to this approach, the transverse momenta (virtualities) of both partons entering the hard process are taken into account and the sum of transverse momenta of the final $c$ and $\bar c$ no longer cancels. Then the differential cross section at the tree-level for the $c \bar c$-pair production reads:
\begin{eqnarray}\label{LO_kt-factorization} 
\frac{d \sigma(p p \to c \bar c \, X)}{d y_1 d y_2 d^2p_{1,t} d^2p_{2,t}} &=&
\int \frac{d^2 k_{1,t}}{\pi} \frac{d^2 k_{2,t}}{\pi}
\frac{1}{16 \pi^2 (x_1 x_2 s)^2} \; \overline{ | {\cal M}^{\mathrm{off-shell}}_{g^* g^* \to c \bar c} |^2}
 \\  
&& \times  \; \delta^{2} \left( \vec{k}_{1,t} + \vec{k}_{2,t} 
                 - \vec{p}_{1,t} - \vec{p}_{2,t} \right) \;
{\cal F}_g(x_1,k_{1,t}^2) \; {\cal F}_g(x_2,k_{2,t}^2) \; \nonumber ,   
\end{eqnarray}
where ${\cal F}_g(x_1,k_{1,t}^2)$ and ${\cal F}_g(x_2,k_{2,t}^2)$
are the unintegrated gluon distribution functions (UGDFs) for both colliding hadrons and ${\cal M}^{\mathrm{off-shell}}_{g^* g^* \to c \bar c}$ is the off-shell matrix element for the hard subprocess. The extra integration is over transverse momenta of the initial
partons. We keep exact kinematics from the very beginning and additional hard dynamics coming from transverse momenta of incident partons. Explicit treatment of the transverse part of momenta makes the approach very efficient in studies of correlation observables. The two-dimensional Dirac delta function assures momentum conservation.
The unintegrated (transverse momentum dependent) gluon distributions
must be evaluated at:
\begin{equation}
x_1 = \frac{m_{1,t}}{\sqrt{s}}\exp( y_1) 
     + \frac{m_{2,t}}{\sqrt{s}}\exp( y_2), \;\;\;\;\;\;
x_2 = \frac{m_{1,t}}{\sqrt{s}}\exp(-y_1) 
     + \frac{m_{2,t}}{\sqrt{s}}\exp(-y_2), \nonumber
\end{equation}
where $m_{i,t} = \sqrt{p_{i,t}^2 + m_c^2}$ is the quark/antiquark transverse mass. In the case of charm quark production at the LHC energies, especially in the forward rapidity region, one tests very small gluon longitudinal momentum fractions $x < 10^{-5}$.  

The matrix element squared for off-shell gluons is taken here in the analytic form proposed by Catani, Ciafaloni and Hautmann (CCH) \cite{Catani:1990eg}. It was also checked that the CCH expression is consistent with those presented later in Refs.~\cite{Collins:1991ty,Ball:2001pq} and in the limit of $k_{1,t}^2 \to 0$, $k_{2,t}^2 \to 0$ it converges to the on-shell formula.

The calculation of higher-order corrections in the $k_{T}$-factorization is much more complicated than in the case of collinear approximation.
However, the common statement is that actually in the $k_{t}$-factorization approach with tree-level off-shell matrix elements a dominant part of real higher-order corrections is effectively included. This is due to possible emission of extra soft (and even hard) gluons encoded
in the unintegrated gluon densities. More details of the theoretical formalism adopted here can be found \textit{e.g.} in Ref.~\cite{Maciula:2013wg}. 
  
In the numerical calculation below we apply the
Kimber-Martin-Ryskin (KMR) unintegrated gluon distributions \cite{Kimber:2001sc,Watt:2003mx} that has been found recently to work very well in the case of charm production at the LHC \cite{Maciula:2013wg}.
As discussed also in Ref.~\cite{Maciula:2016kkx} the $k_{T}$-factorization approach with the KMR UGDF gives results consistent with collinear NLO approach.
For the calculation of the KMR distribution we used here up-to-date collinear MMHT2014 gluon PDFs \cite{Harland-Lang:2014zoa}.
For completness of the present studies, we also use the CCFM-based
JH2013 UGDFs \cite{JH2013} that were applied in the same
context in Ref.~\cite{Cheung:2018tvq}.

The renormalization and factorization scales $\mu^2 = \mu_{R}^{2} =
\mu_{F}^{2} = \frac{m^{2}_{1,t} + m^{2}_{2,t}}{2}$ and charm quark mass 
$m_{c} = 1.5$ GeV are used in the present study. The uncertainties related to the choice of these parameters were discussed, \textit{e.g.} in Ref.~\cite{Maciula:2013wg} and will be not repeated here.

Having calculated differential cross section for $c\bar c$-pair
production one can obtain the cross section for $J\!/\!\psi$-meson
within the framework of improved color evaporation model (ICEM)
\cite{MV2016,CV2017}. The $c\bar c \to J\!/\!\psi$ transition can be
formally written as follows:
\begin{eqnarray}
\frac{d\sigma_{J\!/\!\psi}(P_{J\!/\!\psi})}{d^3P_{J\!/\!\psi}} &= F_{J\!/\!\psi} \int_{M_{J\!/\!\psi}}^{2M_D} d^3 P_{c\bar c} \; d M_{c\bar c} \frac{d\sigma_{c\bar c}(M_{c\bar c},P_{c\bar c})}{ d M_{c\bar c} d^3 P_{c\bar c}} \delta^3(\vec{P}_{J\!/\!\psi}-\frac{M_{J\!/\!\psi}}{M_{c\bar c}} \vec{P}_{c \bar c}),
\end{eqnarray}
where  $F_{J\!/\!\psi}$ is the probability of the $c\bar c \to J\!/\!\psi$ transition which is fitted to the experimental data, $M_{J\!/\!\psi}$ (or $M_{D}$) is the mass of $J\!/\!\psi$ (or $D$) meson and $M_{c\bar c}$ is the invariant mass of the $c\bar c$-system.
Using the momentum relation
\begin{equation}
\vec{P}_{J\!/\!\psi}=\frac{M_{J\!/\!\psi}}{M_{c\bar c}} \vec{P}_{c \bar c}, \;\; \mathrm{where} \;\; \vec{P}_{c \bar c} = \vec{p}_{c} + \vec{p}_{\bar c}, 
\end{equation}
one can easily calculate also rapidity of $J\!/\!\psi$-meson.

%In the last stage we include also a suppression factor:
%
%\begin{equation}
%F_{\mathrm{supp}} = \exp(\frac{-p_{\mathrm{diff}}}{\lambda}) , \;\; \mathrm{where} \;\; \vec{p}_{\mathrm{diff}} = |\vec{p}_{c} - \vec{p}_{\bar c}|,
%\end{equation}
%
%to be discussed in the next section, 
%where $\lambda$ is fitted to the experimental data. 

%-----------------------------------
\section{Numerical results}
%-----------------------------------

Let us start our presentation of numerical results with the discussion of open charm meson production.
Here, we concentrate ourselves on the LHCb open charm data at $\sqrt{s} = 7$ TeV  in $pp$-scattering \cite{Aaij:2013mga}.
A good description of these data will be a starting point for a construction of a theoretical framework
for $\Jpsi$-meson production within the color evaporation model at the given collision energy. In both cases, in the numerical calculations here we follow
the $k_{T}$-factorization approach. 

%----------------------------------------------------------------------------
\begin{figure}[!h]
\begin{minipage}{0.47\textwidth}
  \centerline{\includegraphics[width=1.0\textwidth]{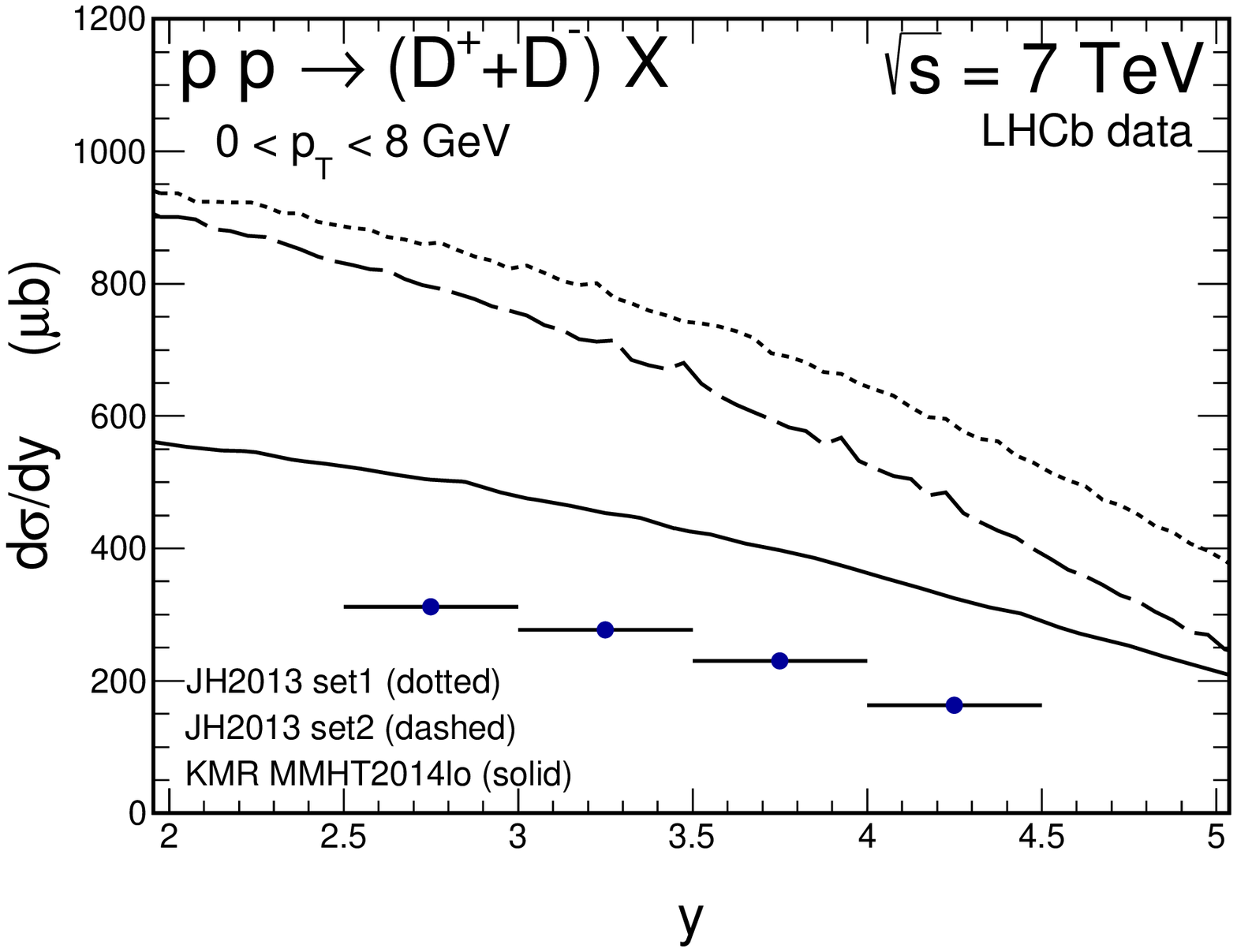}}
\end{minipage}
\begin{minipage}{0.47\textwidth}
  \centerline{\includegraphics[width=1.0\textwidth]{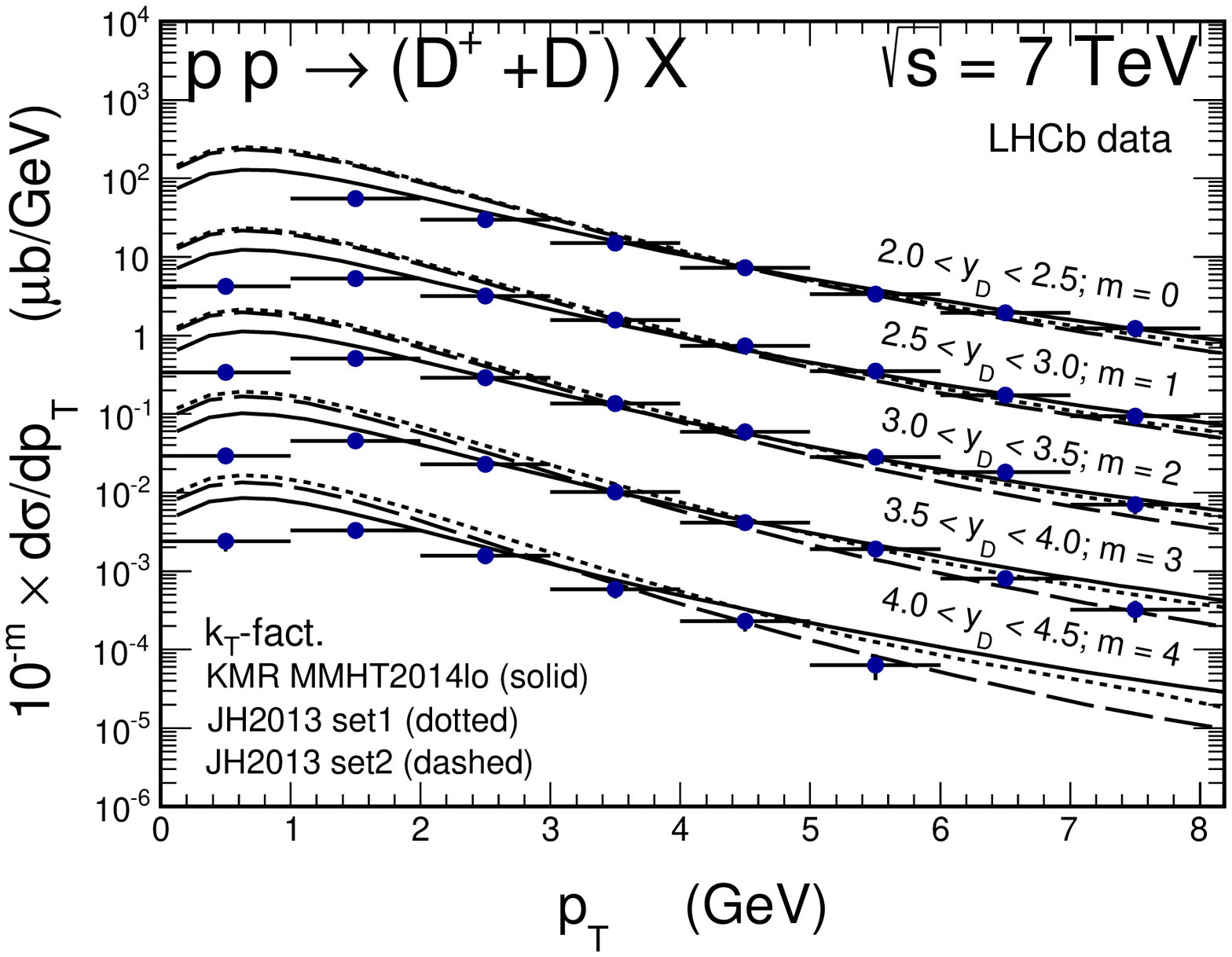}}
\end{minipage}\\
\begin{minipage}{0.47\textwidth}
  \centerline{\includegraphics[width=1.0\textwidth]{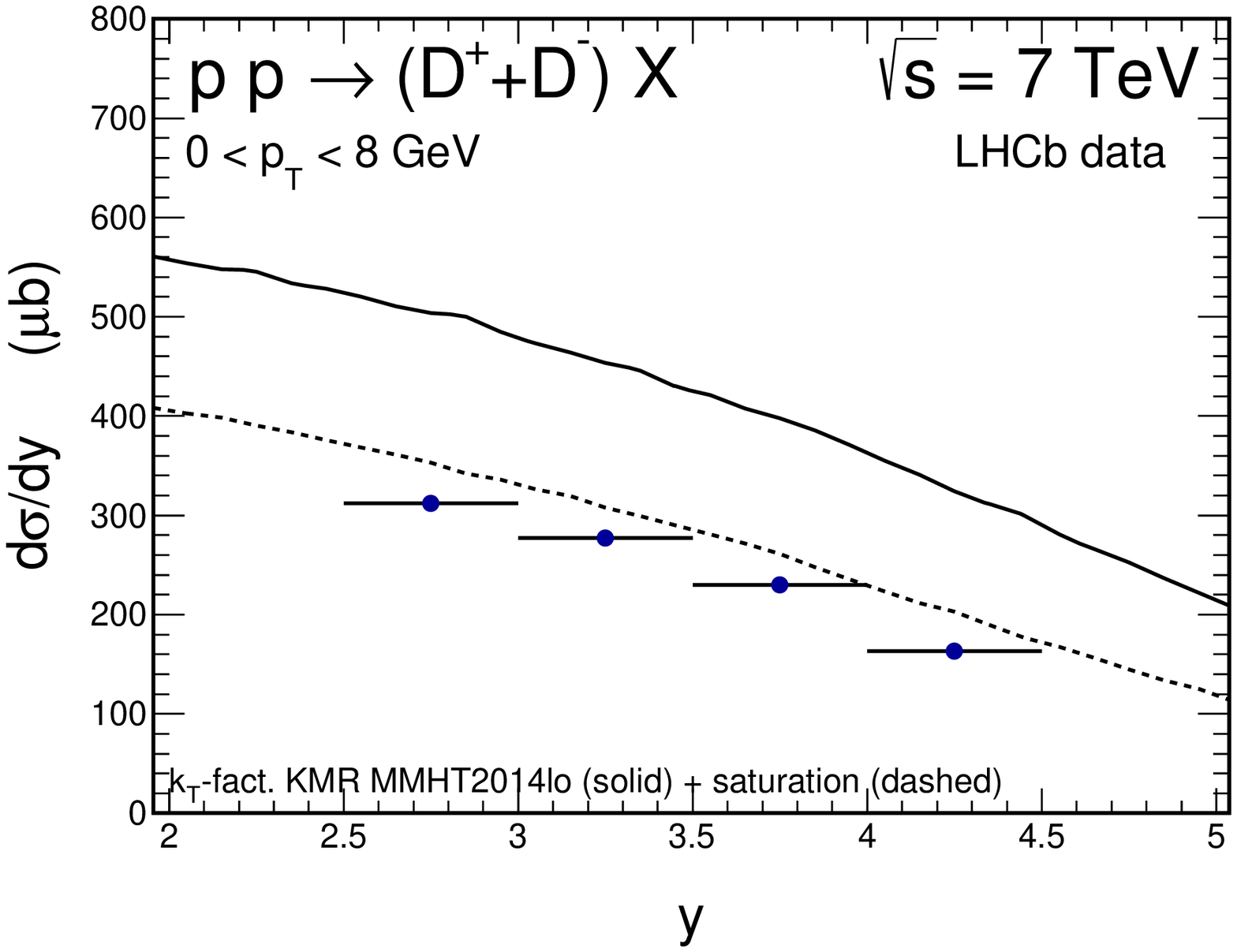}}
\end{minipage}
\begin{minipage}{0.47\textwidth}
  \centerline{\includegraphics[width=1.0\textwidth]{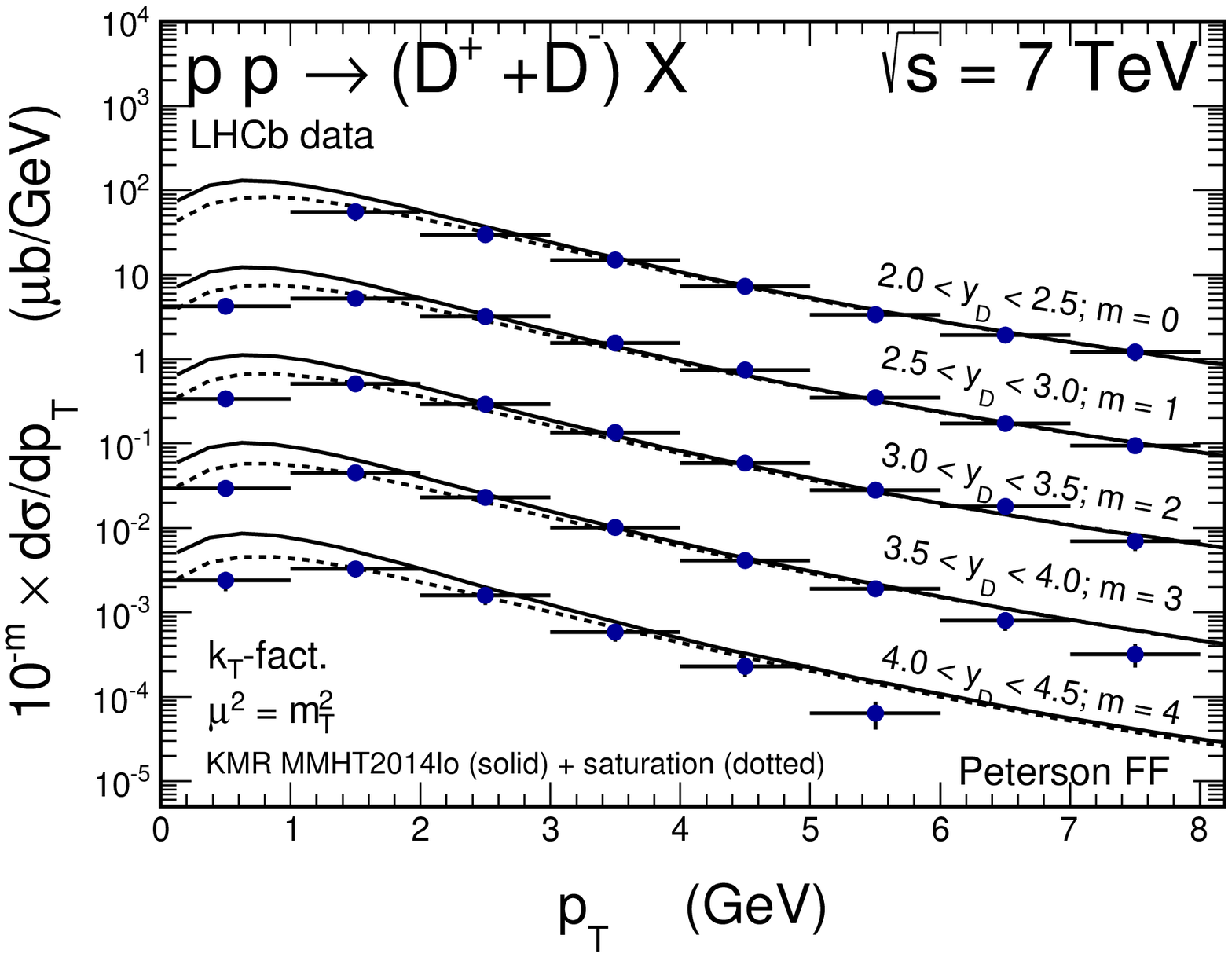}}
\end{minipage}
  \caption{
\small Distributions in rapidity and transverse momentum of $D$ meson
for $\sqrt{s}$ = 7 TeV obtained within the $k_{T}$-factorization realization
of the color evaporation model with the KMR and the JH2013 UGDFs.
}
\label{fig:D_meson}
\end{figure}
%----------------------------------------------------------------------------

In Fig.~\ref{fig:D_meson} we present results of our calculations for charged $D$-meson production in the LHCb kinematical range
of $0 < p_{T} < 8$ GeV and $2 < y < 4.5$. The left and right panels show
the meson rapidity and transverse momentum distributions,
respectively. The $p_{T}$-distributions are shown for different rapidity
bins as specified in the figure. The top panels present the default
$k_{T}$-factorization result for the three different UGDFs: KMR (solid
lines), JH2013 set1 (dotted lines) and JH2013 set2 (dashed lines). Each
of them leads to overestimated rapidity distribution with respect to the
LHCb data. The overestimation is related to small-$p_{T}$ region (see
the right-top panel). The same conclusion is relevant for the two first bins
in transverse momentum. This is the region which is very sensitive to the
small-$k_{T}$ behavior of the UGDF. This uncertain nonperturbative
regime of the UGDFs is still not under theoretical control and can be
treated only in a phenomenological way. Different models follow various
theoretical assumptions and lead to quite different results
\textit{e.g.} for charm production cross sections (see
Ref.~\cite{Maciula:2013wg}). The situation looks much better for
intermediate and larger transverse momenta of $D$ meson, where all of
the applied UGDFs give a very good description of the data points in the
considered $p_{T}$-range. A small discrepancy between the data and the
result of the KMR UGDF appears at larger $p_{T}$'s when moving to the
forward rapidity region. This can be related to a special construction
of the model of UGDF which will be discussed shortly in the end of 
this section. We observe that in this kinematical domain the JH2013 set2
of unintegrated gluon density leads to a better behavior of the predicted
cross section. It might be related to the fact that set2 of the JH2013
UGDF includes the LHC charm data in the fitting procedure, while set1
does not. It is set1 which was used in recent calculation of $\Jpsi$
distributions in Ref.~\cite{Cheung:2018tvq}.    
 
In the bottom panels of Fig.~\ref{fig:D_meson} we illustrate how the description of the LHCb charm data can be improved
within the $k_{T}$-factorization approach by inclusion of the saturation effects in the unintegrated gluon density in proton.
We follow here an useful and pragmatic prescription of these effects as presented in Ref.~\cite{CS2018}. We correct the default KMR unintegrated gluon distribution function
by assuming its saturation as follows: ${\cal F}_{g}(x,k_{t}^2) = {\cal
  F}_{g}(x,Q_{s}^2)$ for small initial gluon transverse momenta
$k_{t}^{2} < Q_{s}^{2}$, where the saturation scale $Q_{s}^{2}(x) =
Q_{0}^{2} \cdot (x_{0}/x)^{\lambda}$. The three free parameters $Q_{0},
x_{0}$ and $\lambda$ are fitted to the LHCb charm data. 
As a result, we get much better description of the LHCb
data for both, the rapidity and the transverse momentum distributions. The saturation effect is visible in the region of $0 < p_{T} < 2$ GeV and seems to be of similar size in each of the considered bins of meson rapidity.   

Let us now go to the main results of the present paper, \textit{i.e.} to theoretical predictions for $\Jpsi$-meson production
within the improved color evaporation model in the version based on the $k_{T}$-factorization approach. For consistency, in the numerical calculations here we keep all the details and parameters as in the case of $D$-meson discussed above.

%----------------------------------------------------------------------------
\begin{figure}[!h]
\begin{minipage}{0.47\textwidth}
  \centerline{\includegraphics[width=1.0\textwidth]{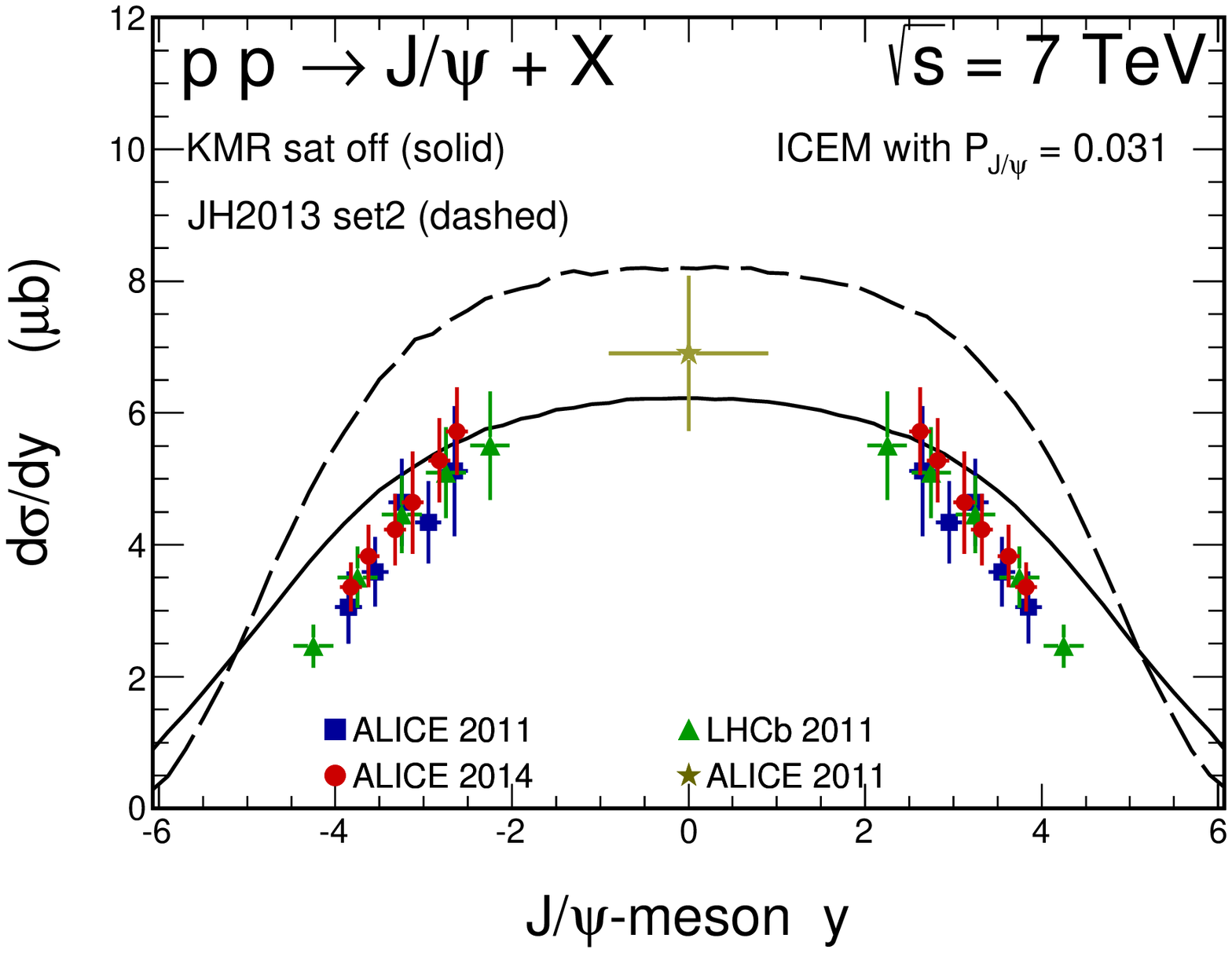}}
\end{minipage}
\begin{minipage}{0.47\textwidth}
  \centerline{\includegraphics[width=1.0\textwidth]{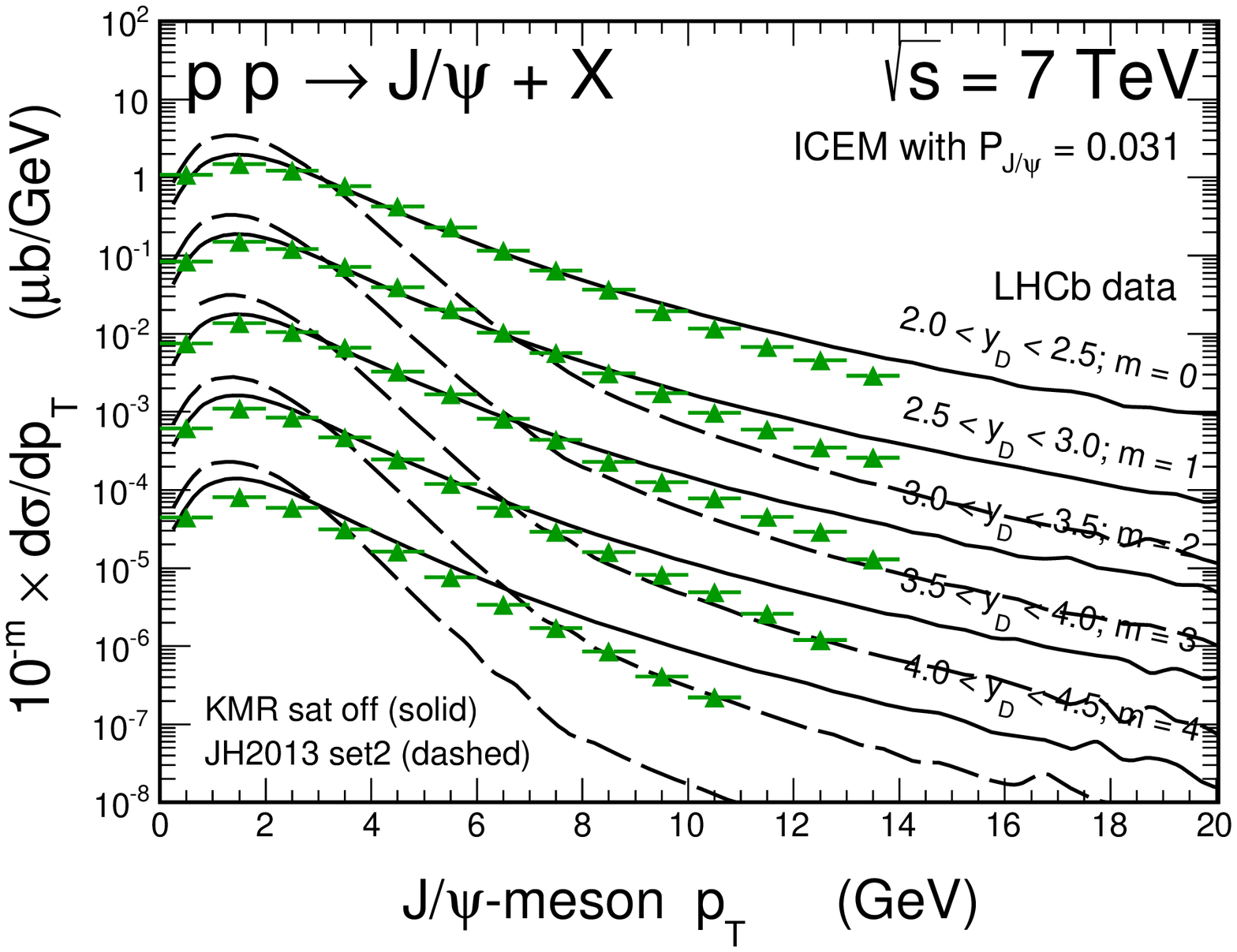}}
\end{minipage}
  \caption{
\small Distributions in rapidity and transverse momentum of $\Jpsi$ meson
for $\sqrt{s}$ = 7 TeV obtained within the $k_{T}$-factorization realization
of the color evaporation model for different UGDFs.
}
\label{fig:JPsi_meson}
\end{figure}
%----------------------------------------------------------------------------

In the left and right panel of Fig.~\ref{fig:JPsi_meson} we show the $\Jpsi$-meson rapidity and transverse momentum distributions, respectively, together with the
ALICE \cite{Aamodt:2011gj,Abelev:2014qha} and the LHCb data \cite{Aaij:2011jh}. Here we present results for the default KMR (solid lines) and for the JH2013 set2 (dashed lines) UGDFs. In this calculation the model parameter $P_{\Jpsi}$ was fixed to 0.031 to describe the ALICE data at midrapidities. Our model rapidity distribution is broader than the experimental one. Similar situation was observed in Ref.~\cite{CS2018} where standard approach for $\Jpsi$-meson production cross section was used. There it was interpreted
as due to gluon saturation at small values of longitudinal momentum
fraction. As already discussed this effect is not explicitly included
when using the default KMR or JH2013 UGDFs. The same idea can also be
the explanation for the color evaporation model approach. There are
visible differences between the predictions of the KMR and the JH2013
set2 UGDFs. In principle, one could fit both rapidity distributions with
the same quality taking different values of $P_{\Jpsi}$. However, the
calculated transverse momentum distributions differs very strongly and
the LHCb data prefers the result with the modified KMR UGDF. The distributions obtained with the JH2013 set2 have completely different $p_{T}$-slope
than the experimental one and falls down much faster. This observation is consistent with the results presented in Ref.~\cite{Cheung:2018tvq}. There, this behavior of the $p_{T}$-distributions was, in our opinion, correctly recognized as a consequence of treatment of the $k_{t} > \mu$ region in the UGDF. The KMR model includes this contribution explicitly. 

\begin{figure}[!h]
\begin{minipage}{0.4\textwidth}
  \centerline{\includegraphics[width=1.0\textwidth]{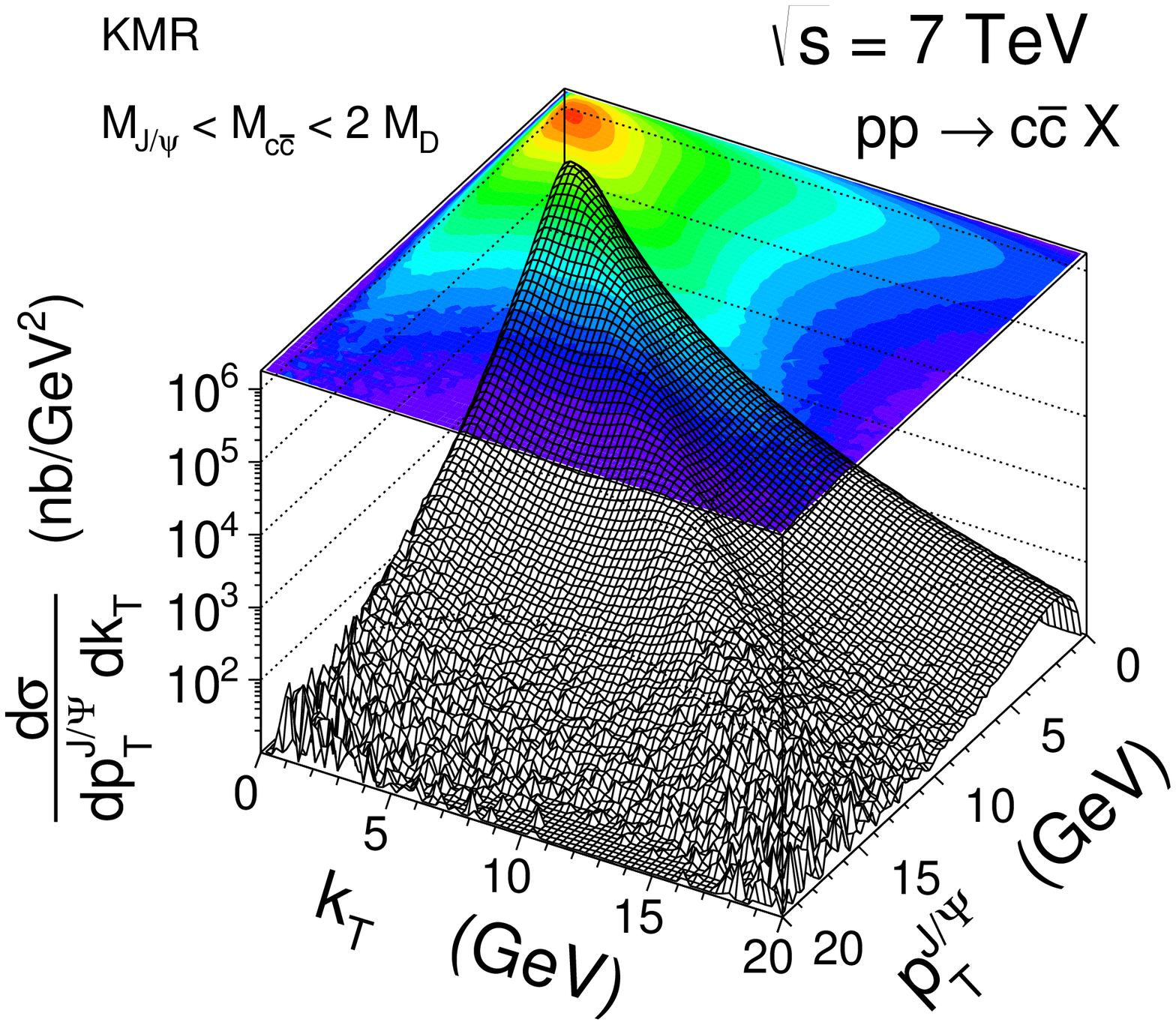}}
\end{minipage}
\begin{minipage}{0.4\textwidth}
  \centerline{\includegraphics[width=1.0\textwidth]{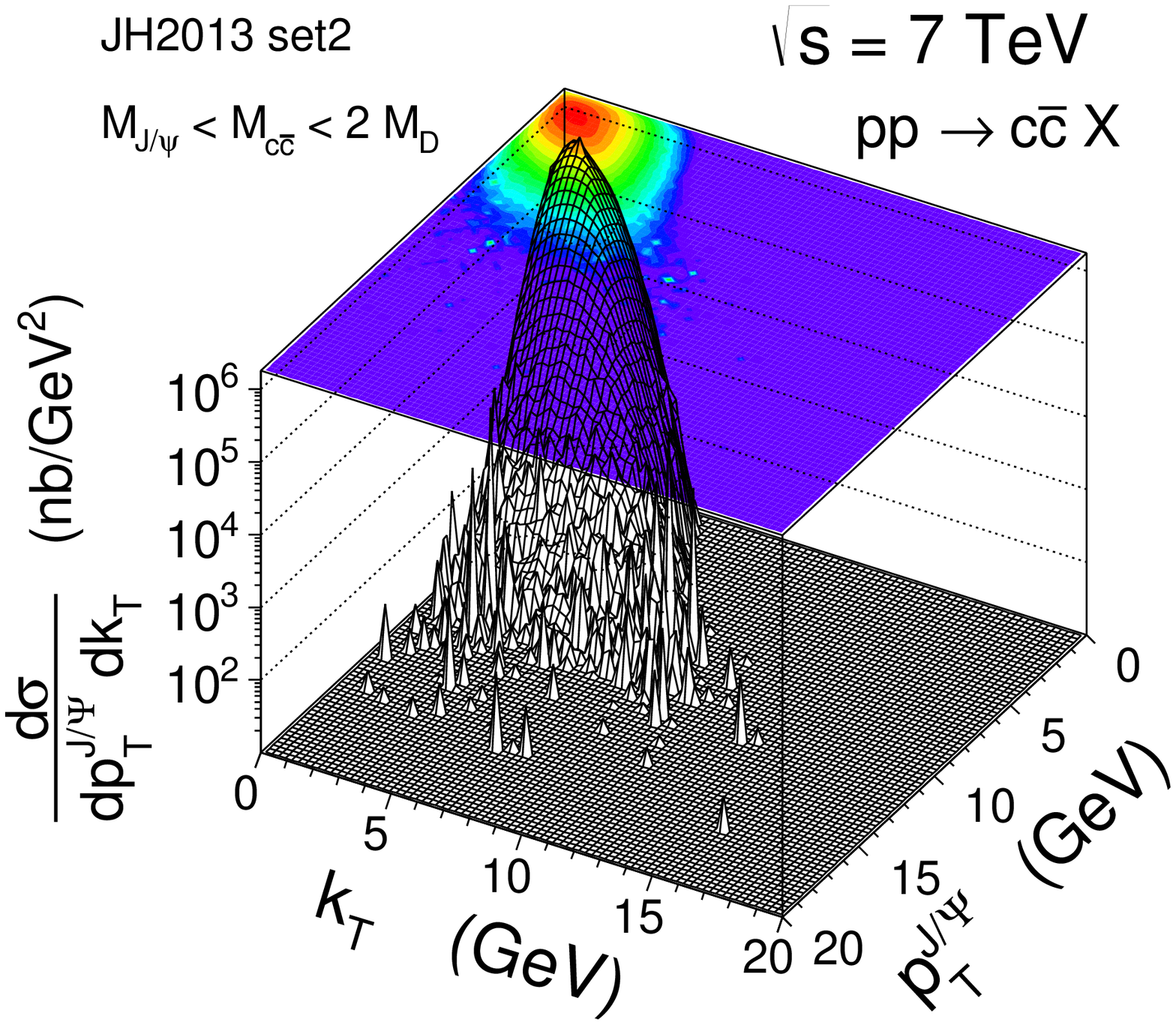}}
\end{minipage}
  \caption{
\small The double-differential cross section as a function of transverse momentum of initial gluon $k_{t}$ and transverse momentum of the meson $p_{T}^{\Jpsi}$ for the KMR (left) and for the JH2013 set2 (right) UGDFs.
}
\label{fig:2}
\end{figure}

It was shown in Ref.~\cite{Maciula:2013wg} that the KMR and CCFM-based UGDFs lead to significant differences in correlation observables for $c\bar c$-pair, \textit{e.g.} in $D\bar{D}$ invariant mass and/or azimuthal angle distributions. In the case of the CCFM-based unintegrated gluon distributions, an improved description of correlation observables
can be obtained once the higher-order process of gluon-splitting is taken into account in an explicit way \cite{Jung:2010ey}. Just to illustrate the differences
we show in Fig.~\ref{fig:2} two-dimensional distributions as a function of transverse momentum of initial gluon $k_{t}$ and transverse momentum of the meson $p_{T}^{\Jpsi}$. The latter observable, within the color evaporation model, in fact represents a modified transverse momentum of the $c\bar c$ -pair. This variable is very strongly correlated with the gluon transverse momentum $k_{t}$. We observe, that the KMR and the JH2013 UGDFs provide very different results. In the case of the KMR, we get long tails in gluon $k_{t}$
that allows for larger $p_{T}^{\Jpsi}$'s (see the left panel). On the
other hand, in the JH2013 model, $k_{t}$ distribution drops down very quickly and therefore production of $\Jpsi$-meson with larger $p_{T}$'s is strongly limited (see the right panel).     

%----------------------------------------------------------------------------
%----------------------------------------------------------------------------

\begin{figure}[!h]

\begin{minipage}{0.47\textwidth}
  \centerline{\includegraphics[width=1.0\textwidth]{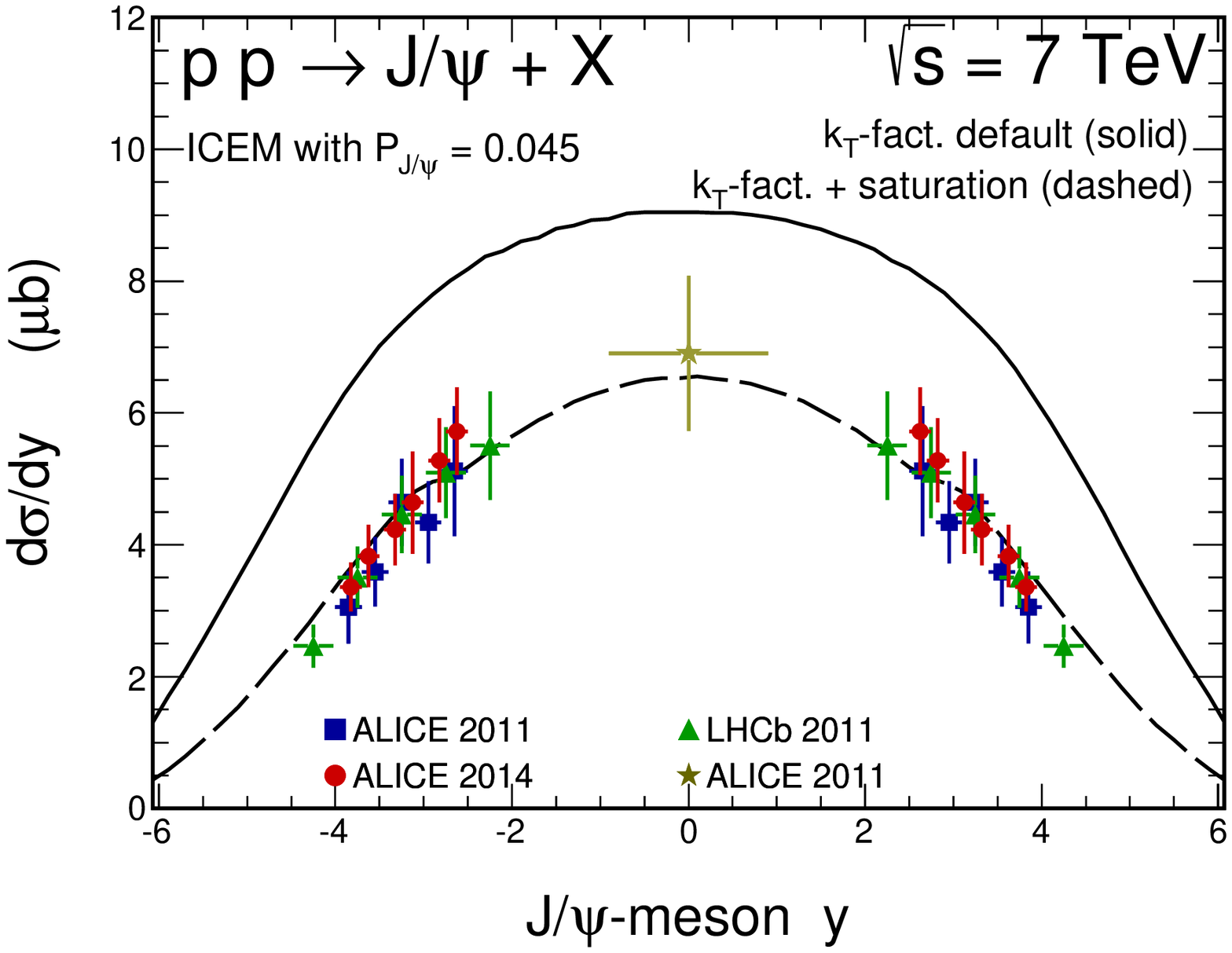}}
\end{minipage}
\begin{minipage}{0.47\textwidth}
  \centerline{\includegraphics[width=1.0\textwidth]{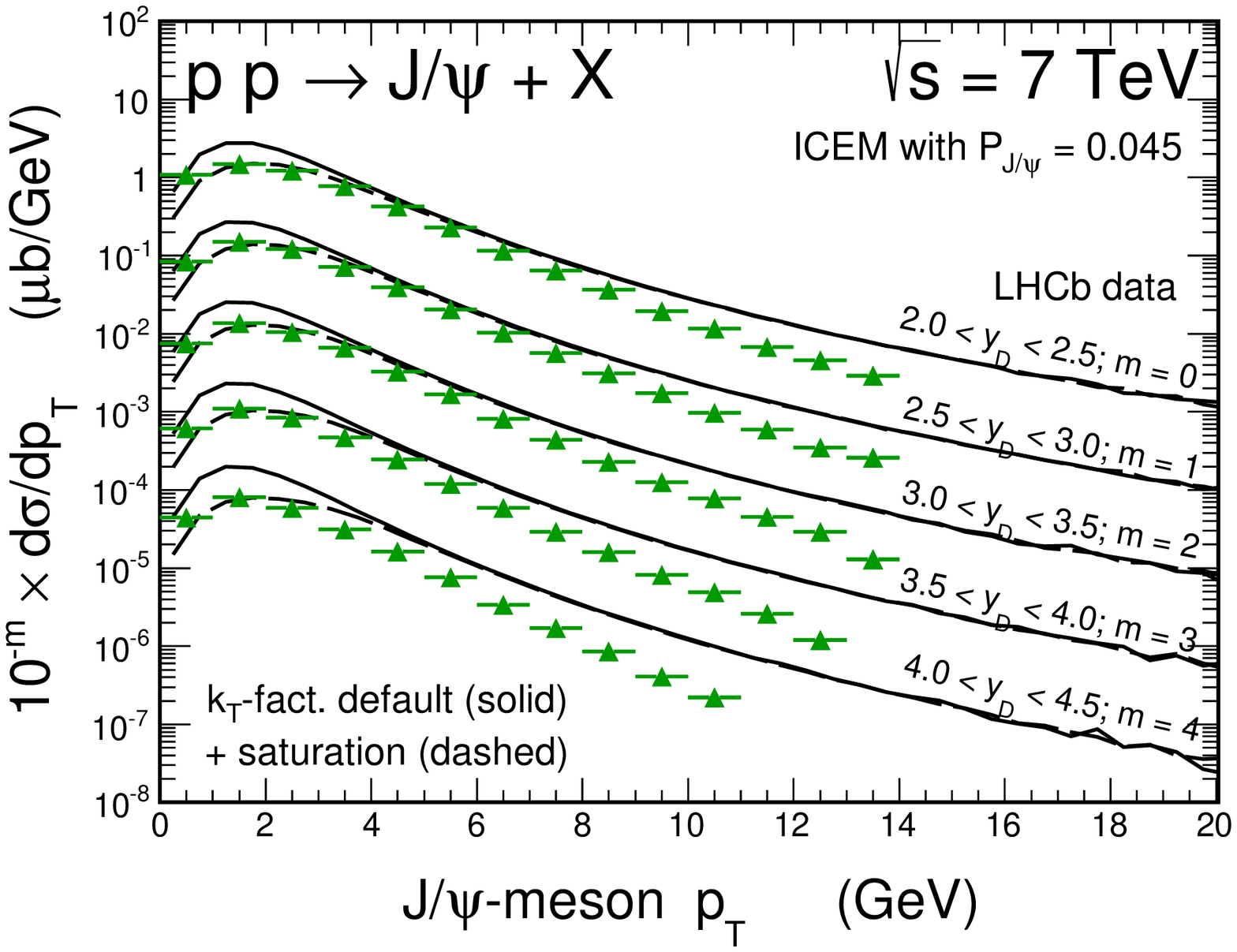}}
\end{minipage}
  \caption{
\small Distributions in rapidity and transverse momentum of $\Jpsi$ meson
for $\sqrt{s}$ = 7 TeV obtained within the $k_{T}$-factorization realization
of the color evaporation model with the saturation effects included.
}
\label{fig:JPsi_meson_sat}
\end{figure}

Now let us take into account the conclusions from the discussion of $D$-meson production from the beginning of this section to
the case of $\Jpsi$-meson production. As we can see from
Fig.~\ref{fig:JPsi_meson_sat} the inclusion of the saturation effects
improves the shape of the rapidity
distribution. Adjusting the normalization via modifying $P_{\Jpsi}$
allows to nicely describe the LHC data. It also improves the shape of the transverse momentum distributions in the region $0 < p_{T}^{\Jpsi} < 4$ GeV but still does not lead to a perfect description of the data points at larger-$p_{T}$'s,
especially in the region of the forward rapidities.

%----------------------------------------------------------------------------
%----------------------------------------------------------------------------

\begin{figure}[!h]
\begin{minipage}{0.47\textwidth}
  \centerline{\includegraphics[width=1.0\textwidth]{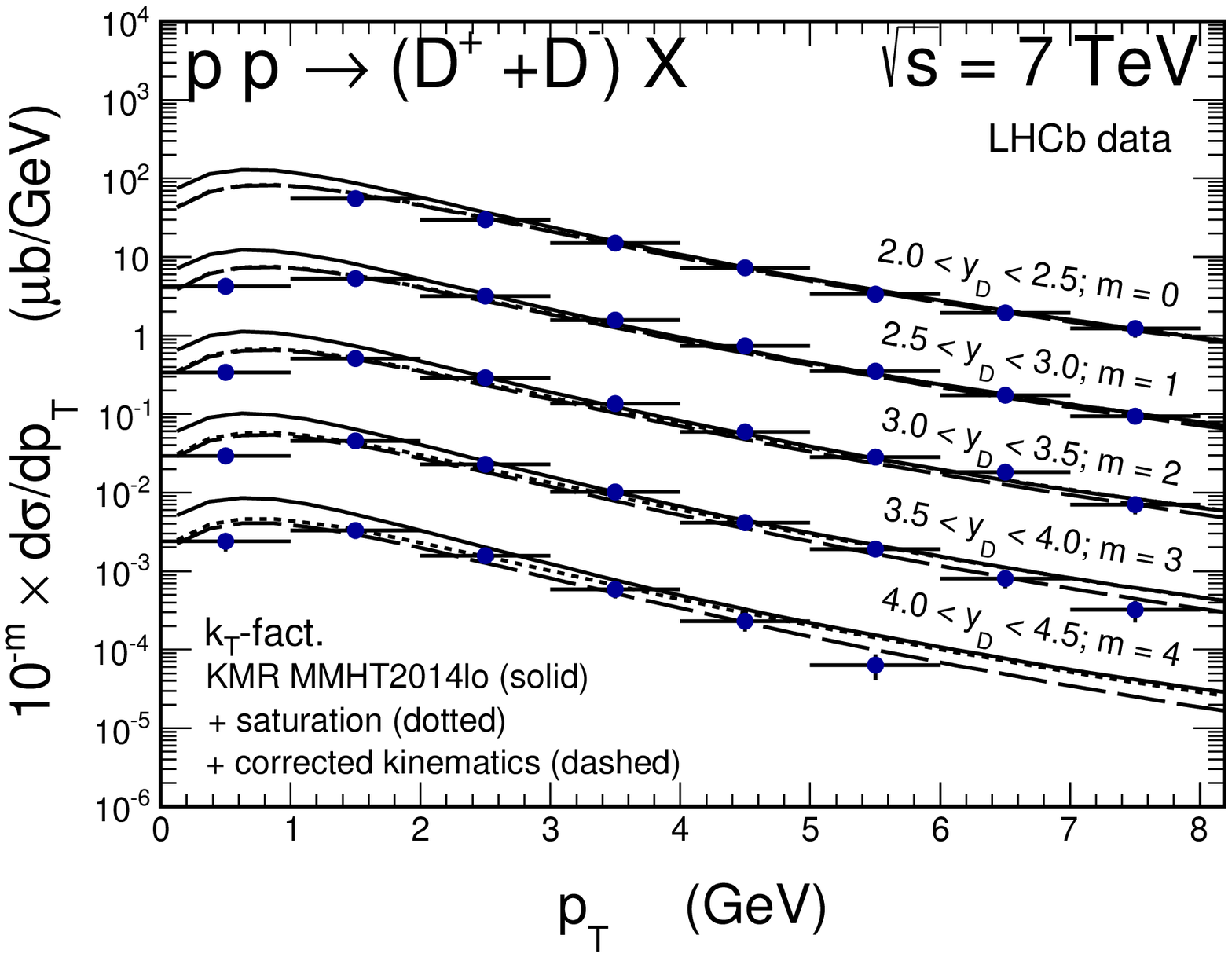}}
\end{minipage}
\begin{minipage}{0.47\textwidth}
  \centerline{\includegraphics[width=1.0\textwidth]{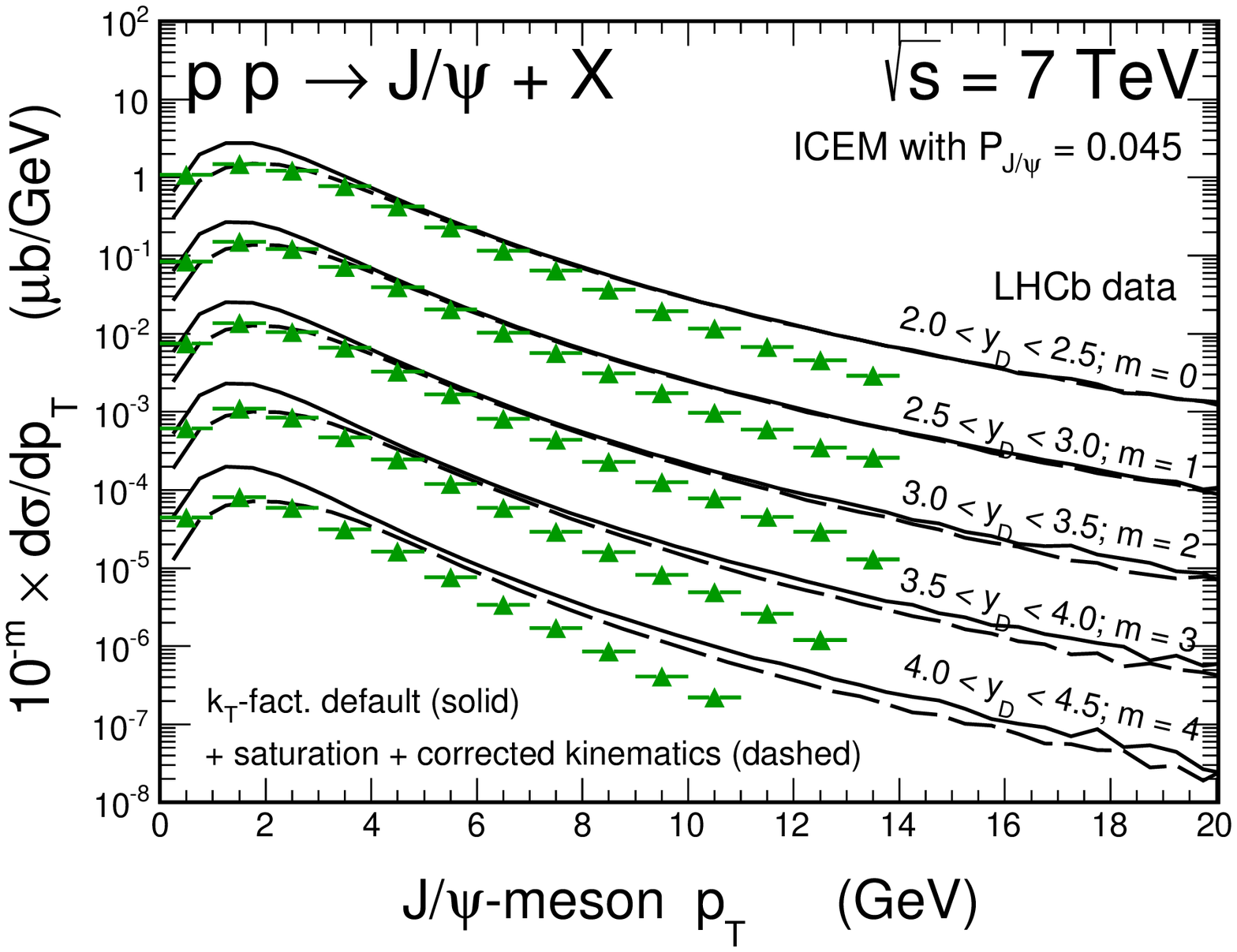}}
\end{minipage}
  \caption{
\small Transverse momentum distribution of $D$ (left) and $\Jpsi$ (right) mesons  for the LHCb experiment for $\sqrt{s} = 7$ TeV. The standard $k_{T}$-factorization calculations with the KMR UGFDF (solid) are shown together with the results of the calculations with the special condition for longitudinal momentum fractions.
}
\label{fig:corrkin}
\end{figure}

Finally, we want to shortly discuss our idea how one could try to
improve the shapes of the $p_{T}$-distributions of $D$-meson 
(and in consequence also of $\Jpsi$-meson) at larger rapidities. 
We propose that one should include in the kinematics the fact that with
the KMR UGDF one do has an additional hidden hard emissions (jet or
two-jets) by its construction.
What we suggest is rather pragmatic: we add into the calculations a special condition $x_{1} + x_{2} + \tilde{x}_{1} + \tilde{x}_{2} < 1$,
where $\tilde{x}_{1} = \frac{k_{1T}}{\sqrt{s}}\exp(y_{\mathrm{max}}+\delta y) + \frac{k_{2T}}{\sqrt{s}}\exp(y_{\mathrm{min}}-\delta y)$ and
$\tilde{x}_{2} = \frac{k_{1T}}{\sqrt{s}}\exp(-y_{\mathrm{max}}-\delta y) + \frac{k_{2T}}{\sqrt{s}}\exp(-y_{\mathrm{min}}+\delta y)$.
Above $y_{\mathrm{min}}= \mathrm{min}(y_{1},y_{2})$ and
$y_{\mathrm{max}}= \mathrm{max}(y_{1},y_{2})$ where $y_{1}$ and $y_{2}$
are rpidities of $c$ and $\bar c$, respectively.
The free parameter $\delta y$ is fitted to the LHCb open charm data. It provides a separation in the rapidity space between the central system and additional hard emissions from the UGDF. Unfortunately, this parameter is energy-dependent. Correcting effectively $x$-values for extra hidden emissions,
not taken explicitly in the $k_{T}$-factorization, improves the shapes 
a bit. It works in the way as in the LHCb data 
(see Fig.~\ref{fig:corrkin}) but it looks that this correction is, 
however, not sufficient. We need a bigger effect for the $\Jpsi$-meson 
than in the case of open charm (please, compare the left and the right panel).

%----------------------------------------------------------------------------
%----------------------------------------------------------------------------

%----------------------------------------------------------------------------
%\begin{figure}[!h]
%\begin{minipage}{0.47\textwidth}
%  \centerline{\includegraphics[width=1.0\textwidth]{dsig_dy_exp_JPsi_7TeV_iCEM.eps}}
%\end{minipage}
%\begin{minipage}{0.47\textwidth}
%  \centerline{\includegraphics[width=1.0\textwidth]{dsig_dpT_exp_JPsi_7TeV_iCEM.eps}}
%\end{minipage}
%  \caption{
%\small Distribution in rapidity and transverse momentum of $\Jpsi$ 
%for KMR UGDF modified for saturation effects.
%}
%\label{fig:3}
%\end{figure}
%----------------------------------------------------------------------------

%Correcting effectively $x$ values for extra hidden emissions,
%not taken explicitly in the $k_{T}$-factorization, improves the shapes
%a bit. It looks that this correction is, however, not sufficient.

%----------------------------------------------------------------------------
%\begin{figure}[!h]
%\begin{minipage}{0.47\textwidth}
%  \centerline{\includegraphics[width=1.0\textwidth]{dsig_dpT_exp_JPsi_7TeV_iCEM_sat_corrkin.eps}}
%\end{minipage}
%  \caption{
%\small Transverse momentum distribution of $\Jpsi$ mesons in the
%improved CEM. Both corrections for saturation and for hidden (mini)jet
%emissions were included (dashed line).
%}
%\label{fig:4}
%\end{figure}
%----------------------------------------------------------------------------

%--------------------------
\section{Conclusions}
%--------------------------

In the present paper we have discussed how to extend color evaporation
model for production of $\Jpsi$ meson to be used in the framework of 
$k_{T}$-factorization approach for production of $c$ and $\bar c$ pairs.
The same was done independently very recently in \cite{Cheung:2018tvq}.
We have included recent developments proposed recently in the literature.
In our calculations we have used the KMR unintegrated gluon 
distributions which allows to describe the single $D$-meson
distributions as well as meson correlation observables. 

Rapidity and transverse momentum distributions of $\Jpsi$ mesons
have been calculated and the normalization factors, being a probability
of $c \bar c$ soft transition to color singlet $S$ wave quarkonium,
have been obtained. We could not describe all the world data
with exactly the same normalization.

Inspired by our earlier work \cite{CS2018} we have tried to include saturation
effects that modify the KMR UGDF for small values of longitudinal momentum
fractions. Including such an effect improves agreement of the $\Jpsi$
distributions as well as those for $D$ meson production.

A still better agreement for the transverse momentum distributions  can be 
achieved by including explicitly hidden, in the $k_{T}$-factorization
with the KMR UGDF, emissions in calculating $x$ values for UGDFs.

In summary, a reasonable description of the data for $\Jpsi$ can be 
achieved in the color evaporation model supplemented for
the $k_{T}$-factorization approach for production of $c \bar c$ pairs.
Simultaneously a rather good description of the $D$ meson production is
achieved with the same UGDF.

A relatively good description obtained within the improved color 
evaporation model does not proof that we have understood 
the underlying mechanism.
A similar quality of agreement with experimental data can be obtained 
within the $k_{T}$-factorization npQCD approach applied directly to 
$\Jpsi$ production \cite{CS2018}.
Production of pairs of quarkonia may provide further tests.

%We have discussed how the average momentum of quark/antiquark in
%the $c \bar c$ rest frame changes from mid to forward rapidities.
%We have introduced a special form factor, with one new parameter,
%which can take into account the quarkonium wave function effects.

%With the one extra parameter we have been able to describe the LHC data
%with surprisingly good precision.

\vspace{1cm}

{\bf Acknowledgments}

This study was partially supported by the Polish National Science Center
grant DEC-2014/15/B/ST2/02528 and by the Center for Innovation and
Transfer of Natural Sciences and Engineering Knowledge in Rzesz{\'o}w.

%-------------------------------------------------------------------------------------

\end{document}